\documentclass[12pt]{article}
\usepackage{a4}
\usepackage{epsf}

\newcommand{\beq}{\begin{equation}}
\newcommand{\eeq}{\end{equation}}
\newcommand{\bea}{\begin{eqnarray}}
\newcommand{\eea}{\end{eqnarray}}
\newcommand{\nn}  {\nonumber}

\def\mom{{_{\widetilde{\rm \kern -1pt M\kern -1pt O\kern -1pt M}}}}
\def\momt{$\widetilde{\rm\scriptstyle M\kern -1ptO\kern -1ptM}$ }
\def\ms{{_{\kern 1pt\overline{ \rm M \kern -1ptS}}} }
\def\mst{${\rm\scriptstyle \overline{ M\kern -1pt S}}$ }

\begin{document}

\title{Three-loop beta function and 
non-perturbative $\alpha_s$ in 
asymmetric momentum scheme}

\author{ Ph. Boucaud$^a$,  J.P. Leroy$^a$, 
J. Micheli$^a$, \\  O. P\`ene$^a$  and C. Roiesnel$^b$ }
 \par \maketitle
\begin{center}
{\it $^a$Laboratoire de Physique Th\'eorique et Hautes
\'Energies\footnote{Laboratoire
associ\'e au
Centre National de la Recherche Scientifique - URA D00063}
\\
{Universit\'e de Paris XI, B\^atiment 210, 91405 Orsay Cedex,
France}\\$^b$ Centre de Physique Th\'eorique\footnote{
Unit\'e Mixte de Recherche C7644 du Centre National de 
la Recherche Scientifique\\ 
\\e-mail: Philippe.Boucaud@th.u-psud.fr, roiesnel@cpht.polytechnique.fr
}  de l'\'Ecole Polytechnique\\
91128 Palaiseau Cedex, France 
}
\end{center}

\begin{abstract}
We determine 
the three-loop coefficient of the beta 
function  in the asymmetric momentum subtraction 
scheme in Landau gauge.  This scheme is convenient for lattice studies
of $\alpha_s$, the running coupling constant of QCD.
We present high statistics lattice results for $\alpha_s$
in the SU(3) Yang-Mills theory without quark,
compare with  the three-loop running and 
extract the value of the corresponding $\Lambda_\ms$ parameter.
We estimate the systematic error coming from four-loop terms.
We obtain the result~:
$ \Lambda_\ms= 295\, (5) (15)
 \frac {a^{-1}(\beta=6.0)}{1.97 {\rm GeV}}\,  {\rm MeV}$.

\end{abstract}
\begin{flushright} LPTHE Orsay-98/65\\  hep-ph/9810437
\end{flushright}

\newpage

Momentum subtraction schemes are interesting
because they provide  regularization independent
renormalization methods which can be used non-perturba\-ti\-vely, contrary
to the popular \mst schemes based on the dimensional regularization.
Realized on the lattice, momentum subtraction schemes eliminate
 the need for lattice perturbation theory
which has proved to be complicated and not very accurate. Usefulness
of these schemes is exemplified  in the
non-perturbative renormalization of operators \cite{guido} 
which becomes widely used by now 
 in the lattice approach.

In this paper we determine the three-loop running in an asymmetric momentum
subtraction scheme, compare with our high statistics numerical results
for $\alpha_s$ and extract the value for  the $\Lambda_\ms$ parameter
in the SU(3) Yang-Mills theory without quark.
In  section 1, we 
recall the definitions of the \momt renormalization scheme and 
define $g_\mom$, a coupling constant convenient for lattice studies \cite{alles,beta2}.
In section 2, we give the relation between this coupling constant and some 
quantities computed  in the \mst scheme in ref.\cite{davyd}.
In section 3, we extract the  three-loop coefficient of the beta function 
in the \momt scheme and in section 4, we present our results for the 
coupling constant and the $\Lambda_\ms$ parameter.

\section{Definition of the coupling constant in the \momt scheme}
\label{definition}

\noindent
In this section, we suppose that we are able to compute or measure
in some way the  two- and three-point gluonic Green functions
and define a convenient coupling constant.

\noindent
When one of the momenta is equal to zero, the Lorentz 
structure of the three-point Green function, 
$G^{(3)abc}_{\mu\nu\rho} (p,-p,0)$, can depend a priori
on four tensors  : 
$\delta_{\mu\nu} p_\rho$, $\delta_{\mu\rho} p_\nu$,
$\delta_{\nu\rho} p_\mu$ and $p_\mu p_\nu p_\rho$. 
In    Landau gauge, only one possibility is left 
by the transversality conditions and we can write~:
\bea
G^{(3)abc}_{\mu\nu\rho} (p,-p,0) \ &=& \  2 \ f^{abc}
\left(\delta_{\mu\nu} - {p_{\mu} p_{\nu} \over p^2}\right) p_\rho \ G^{(3)}(p^2)
\label{g3}\eea
for the bare Green function.
Note that only the completely antisymmetric group structure functions $f^{abc}$
can appear in this expression :  the Bose symmetry of the three point function,
i.e. the symmetry over  $p \leftrightarrow -p $, $ \mu \leftrightarrow \nu$
and $a \leftrightarrow b$, combined with the
oddness of the Lorentz tensor when $p \leftrightarrow -p $ 
forbids terms with the symmetric color structure $d^{abc}$.

\noindent
The bare two-point Green function in Landau gauge writes~:
\bea
G^{(2)ab}_{\mu\nu} (p,-p) \ &=& \ \delta^{ab}
\left(\delta_{\mu\nu} - {p_{\mu} p_{\nu} \over p^2}\right)\  G^{(2)}(p^2)\, , \nn \\
G^{(2)ab}_{\mu\nu} (0,0) \ &=& \ \delta^{ab} \delta_{\mu\nu}\  G^{(2)}(0) \, .
\label{g2}\eea

\noindent
The scalars functions $G^{(3)}$ and $G^{(2)}$ can be extracted from the
Green functions with~:
\bea
G^{(3)}(p^2) &=& {1 \over 144 \ p^2} \ f^{abc} \delta^{\mu\nu} \ p^\rho 
               \ G^{(3)abc}_{\mu\nu\rho} (p,-p,0) 
                  \, ,\nn\\		 
G^{(2)}(p^2) &=& {1\over 24} \sum_{a,\mu}\  G^{(2)aa}_{\mu\mu} (p,-p)\, ,\nn\\
G^{(2)}(0)   &=& {1\over 32} \sum_{a,\mu} \ G^{(2)aa}_{\mu\mu} (0,0)\, .
\eea

\noindent
The wave function renormalization for the gluon, $Z_\mom$,  and the renormalized
coupling constant in the \momt scheme are then defined by:

\bea
Z_\mom(\mu^2)  &=& p^2 \ G^{(2)}(p^2) \vert_{_{p^2 = \mu^2}}\label{Z}\, , \\
\nn\\
g_\mom(\mu^2)  &=& \left. {G^{(3)}(p^2) \ Z_\mom^{3/2}(\mu^2) 
                    \over G^{(2)}(p^2) G^{(2)}(p^2) G^{(2)}(0)}
		    \right|_{_{p^2 = \mu^2}}\, ,
		    \label{g}
\eea
where $\mu$ is the renormalization scale. The interpretation of eqs. (\ref{Z}-\ref{g})
is standard~: the momentum scheme fixes the renormalization constants so that the
two- and three-point functions take their tree values with the substitution of the
bare coupling by the renormalized one. In the usual
${\rm\scriptstyle M\kern -1ptO\kern -1ptM}$ 
 scheme, one chooses
to work  at the symmetric Euclidean point $p_1^2 = p_2^2 = p_3^2 = \mu^2$.
Here the \momt scheme is defined from an asymmetric point, when 
one of the momenta is equal to zero.

\section{Connection with  \mst calculations}
\label{connection}

In ref. \cite{davyd}, the two- and three-point Green functions have been 
computed at two-loop 
with the dimensional regularization in an arbitrary covariant gauge. They define
the (bare) form factors from the (bare) vertex function:
\bea
\Gamma_{\mu_1\mu_2\mu_3}^{abc}(p,-p,0) \ = \ f^{abc} \Big\{ ( 
2 \delta_{\mu_1\mu_2} p_{\mu_3} &-& \delta_{\mu_1\mu_3} p_{\mu_2} 
                           - \delta_{\mu_2\mu_3} p_{\mu_1} ) \ T_1(p^2)\nn\\
&-& p_{\mu_3} \left( \delta_{\mu_1\mu_2} 
    - {p_{\mu_1} p_{\mu_2} \over p^2} \right)\  T_2(p^2)\Big\}\, .
\eea
At zeroth order in perturbation theory, $T_1 = 1$ and $T_2 = 0$.

\noindent
If we restrict to Landau gauge and insert these form factors into 
the definition (\ref{g}) for the coupling constant in the \momt scheme, 
 we get~: 
\bea 
g_\mom(\mu^2) \ = \ 
\left( {\mu^2 {\rm e}^\gamma \over 4 \pi} \right)^{-\epsilon/2}
 g_0 \left.\left( T_1 - {1\over 2} T_2 \right)\right|_{_{p^2 = \mu^2}}  
 Z_\mom^{3/2} \label{gmombare}\, ,
 \eea
where $g_0$ is the bare coupling constant and 
a common scale, $\mu^2$,  has been chosen 
for the renormalization point used to define 
$g_\mom$ and for the dimensional regularization.

\noindent
To renormalize in the \mst scheme, renormalization constants 
and a renormalized coupling
constant have to be introduced. The definitions are:
\bea
\Gamma_{\ms}^{\mu_1\mu_2\mu_3}(p,-p,0) &=& 
               Z_1 \ \Gamma^{\mu_1\mu_2\mu_3}(p,-p,0)\label{gammaMS}\, ,\\
G^{(2)}_{\ms}(p^2) &=& Z_3^{-1} G^{(2)}(p^2)\label{propMS}\, ,\\
g_0 &=&  \left( {\mu^2 {\rm e}^\gamma \over 4 \pi} \right)^{\epsilon/2}
         \ g_\ms \ Z_1 Z_3^{-3/2}\, ,\label{g0gms}
\eea 
where the suffix \mst indicates renormalized quantities in the \mst scheme.
$Z_1$ and $Z_3$ are the renormalization constants in the \mst scheme;  
they are defined in the usual way to remove  the $1/\epsilon^j$ singular terms
present in the bare quantities.
From (\ref{gammaMS}) we get $ T_{1\ms} = Z_1 T_1$ and   $ T_{2\ms} = Z_1 T_2$
and from
the definitions (\ref{Z}) and (\ref{propMS})~: 
\bea
p^2 G^{(2)}_{\ms}(p^2)\Big|_{_{p^2 = \mu^2}} 
\ =\ Z_3^{-1} p^2 G^{(2)}(p^2)\Big|_{_{p^2 = \mu^2}} 
\ =\ {Z_\mom\over Z_3}\, .
\eea
Collecting everything, the relation between the \momt coupling constant  
  and  \mst  quantities  
is:
\bea
\alpha_\mom(\mu^2)\ 
&=& \ \alpha_\ms(\mu^2)
              \left.\left( T_{1\ms} - {1\over 2} T_{2\ms}\right)^2 \right|_{p^2=\mu^2} 
	      \left( \mu^2 G^{(2)}_{\ms}(\mu^2)\right)^3\, ,	\label{relation}      
\eea
where $\alpha_\mom \equiv {g_\mom^2 / 4 \pi}$ and 
$\alpha_\ms \equiv {g_\ms^2 / 4 \pi}$.

\section{The three-loop $\beta$ function in the \momt scheme}

The $\beta$ function in the \momt and \mst schemes are defined by~:
\bea
\beta_\mom(\alpha_\mom)   = 
 \mu\frac{\partial \alpha_\mom}{\partial \mu} &=&
 -\frac{\tilde\beta_0}{2\pi}\alpha_\mom^2
 -\frac{\tilde\beta_1}{4\pi^2}\alpha_\mom^3 
 -\frac{\tilde\beta_2}{64\pi^3}\alpha_\mom^4
 -\frac{\tilde\beta_3}{128\pi^4}\alpha_\mom^5
 -...\nn\\
\beta_\ms(\alpha_\ms)  \ =\  
 \mu\frac{\partial \alpha_\ms}{\partial \mu} &=&
 -\frac{\beta_0}{2\pi}\alpha_\ms^2
 -\frac{\beta_1}{4\pi^2}\alpha_\ms^3 
 -\frac{\beta_2}{64\pi^3}\alpha_\ms^4
 -...\label{betams}
 \eea

\noindent
From \cite{davyd}, the perturbative expansions for the 
gluon propagator and the form factors 
$T_{1\ms}$ and $T_{2\ms}$ are known up to two loops in any covariant gauge. 
Eq.(\ref{relation}) can then be written as~:
\bea
\alpha_\mom \ = \ \alpha_\ms  
\left( 1 \ +\  a\  \frac{\alpha_\ms} { 4 \pi} 
         \ +\ b\  \frac{\alpha_\ms^2} { 16 \pi^2} + ... \, ,\label{lien}
\right)
\eea
with known coefficient $a$ and $b$. This gives a relation between 
the $\beta$ functions~:
\bea
\beta_\mom(\alpha_\mom)  \ =\ \beta_\ms(\alpha_\ms) 
\left( 1 \ +\   a\  \frac{\alpha_\ms} { 2 \pi} 
         \ +\   3 b\  \frac{\alpha_\ms^2} { 16 \pi^2} + ...\right)\, .\label{aa}
\eea
Using eq.(\ref{lien}) and the definitions (\ref{betams}), we can
expand (\ref{aa}) in power of $\alpha_\ms$  
and identify the terms with the same power; we get~:
\bea
\tilde\beta_0 &=& \beta_0\label{beta0} \, ,\\
\tilde\beta_1 &=& \beta_1\label{beta1}\, ,\\
\tilde\beta_2 &=& \beta_2 + 2 \left( b - a^2 \right) \beta_0 - 4 a \beta_1\, .
\label{beta2}
\eea
Eqs. (\ref{beta0},\ref{beta1}) are   usual results~: 
 the two first coefficients of the $\beta$ function do not 
 depend on the renormalization scheme.
Eq. (\ref{beta2}) tells us that the two-loop results in \mst 
for the gluon propagator, $T_1$ and $T_2$ 
(available in \cite{davyd}) are  sufficient to get the difference
between the three-loop coefficients
of the $\beta$ function in the \momt and the \mst schemes. 
In eq.(\ref{beta2}) $a$, $b$ and $\beta_2$ have to be computed in Landau gauge. 
But it is known  that $\beta_2$ in \mst does not  depend on the gauge \cite{lyrin}
so we  can  use the standard  result  in  Feynman gauge,
 $\beta_2 = 2857  - \frac{5033}{9}\  N_f + \frac{325}{27}\  N_f^2$
 \cite{taras}.

\noindent
Finally, the three-loop $\beta$ function in the \momt scheme
 and Landau gauge is found to be:
\bea
\tilde\beta_0 &=& 11 - \frac{2}{3} N_f \, ,\nn\\
\tilde\beta_1 &=& 51 - \frac{19}{3} N_f \, ,\nn\\
\nn\\
\tilde\beta_2 &=& 
\frac{186747}{32} - \frac{1683}{2}\zeta_3\nn\\
\nn\\
&&
-\left( \frac{35473}{48} - \frac{65}{3}\zeta_3\right) \ N_f 
-\left( \frac{829}{27} - \frac{16}{9}\zeta_3\right) \ N_f^2 
+ {\frac{16}{9}}\ N_f^3 \, ,
\eea
where $\zeta_3 \simeq 1.2020569...$ is the value of the Riemann's zeta function 
and $N_f$ is the number of flavor. In the following we will work  in the flavorless
case, $N_f$ = 0, for which $\tilde\beta_2 \ \simeq\ 4824.31$.

\section{$ \Lambda_\ms$ from $\alpha_\mom$}

The implementation of the  \momt scheme on the lattice is straightforward
\cite{alles,beta2}.
Varying the external momentum, the \momt coupling constant can be obtained
directly   in one simulation for several
values of the scale $\mu^2$.
In ref.\cite{beta2}, the two- and three-point Green functions have been 
measured on the lattice in Landau gauge with high statistics and for several
lattice spacings and volumes in the flavorless  case. For large values
of the momentum, lattice artifacts of $O(a^2 p^2)$ affect the
Green functions. Numerically as expected these effects  are seen to
decrease when $\beta$ increases. 
On the lattice the  gauge fixing algorithm  leads to
the relation $\frac{2}{a} \sin(\frac{a p_\mu}{2}) A_\mu(p) \ = \ 0$ 
while $p_\mu A_\mu(p)$ does not vanish.
And actually  we have seen on our data \cite{beta2} that 
the dominant part of $O(a^2 p^2)$ artifacts in $\alpha_\mom$ 
is corrected by the substitution  of the momenta
$p_\mu$ by the lattice momenta $\frac{2}{a} \sin(\frac{a p_\mu}{2})$
in eqs. (\ref{g3}) and (\ref{g2}). 
This was already noticed in \cite{alles} ; 
in other contexts, authors have shown that this choice for the lattice momentum
was indeed favored by their data, see for example \cite{moment,ape}.

\begin{figure}
\begin{center}
\leavevmode
\vspace*{-4.0cm}
\epsfbox[100 200 510 860]{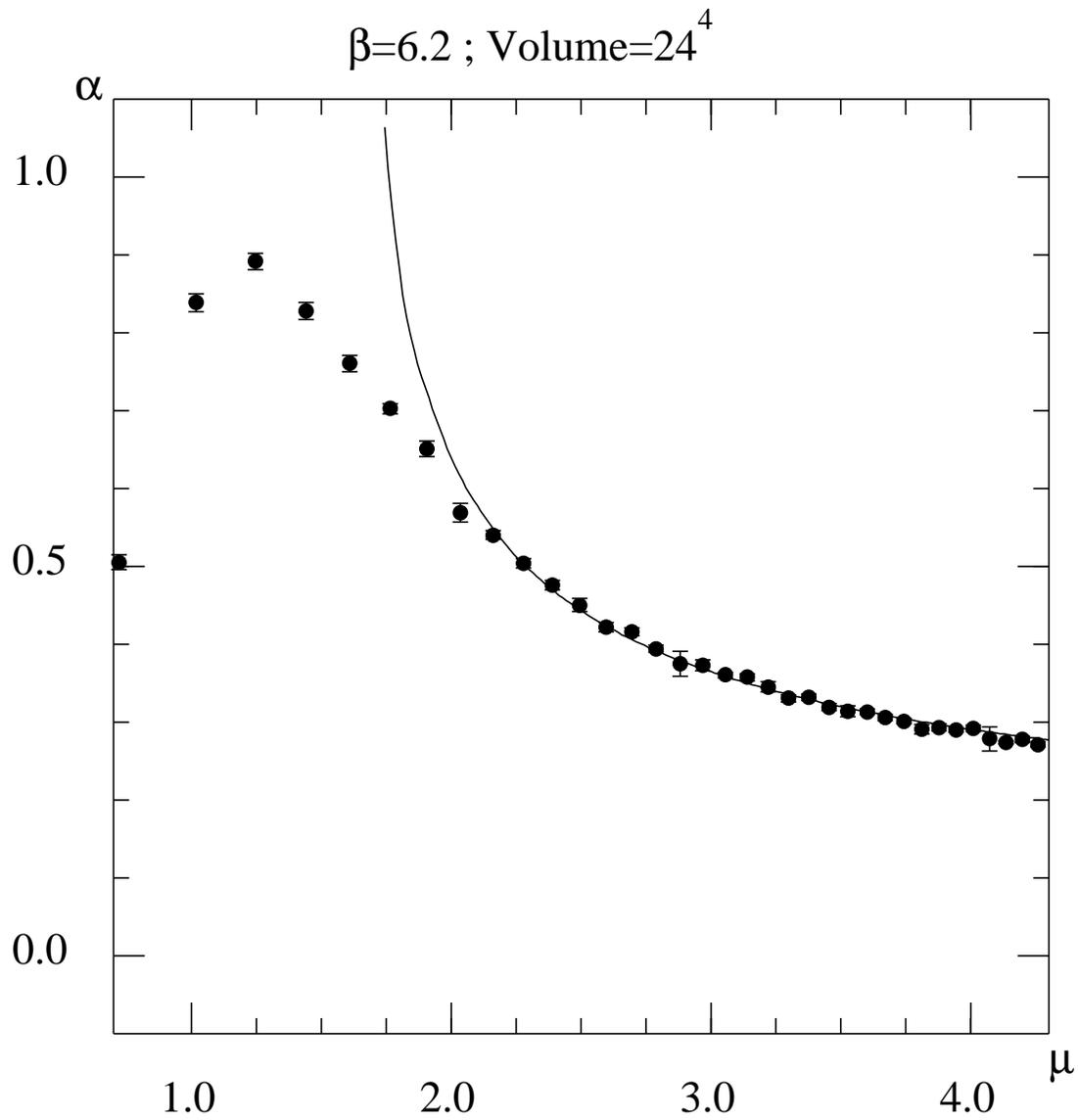}
\caption{ The QCD coupling constant $\alpha_\mom$ as a function of the scale $\mu$
(in GeV). 
The full line is   the three-loop running. }
\label{figure2}
\end{center}
\end{figure}

\noindent
We give now some results from the three-loop analysis, 
see \cite{beta2} for    details on  the lattice settings.
In Fig.\ref{figure2}, we give the behavior of 
the coupling constant, $\alpha_\mom$, as a function of the scale $\mu$ obtained 
from the simulation and compared with the integration of the three-loop 
beta-function. A nice scaling is apparent for scales larger than  2  GeV.
 
 \noindent 
 To calibrate the lattice at $\beta=6.2$, we have used the value for the the lattice
spacing which has  been measured recently  with a non-perturbatively 
improved action (free from {O(a)} artifacts) \cite{ape}. The measured  value is~: 
$a^{-1}(\beta=6.2)\ =\ 2.75(18)$ GeV.  Other lattices have been calibrated
relatively to the one at $\beta = 6.2$ with the results for  
$a \sqrt{\sigma}$, the string tension in lattice units, published in \cite{bali}.
We took: $a^{-1}(\beta=6.0) = 1.97$ GeV, 
$a^{-1}(\beta=6.2) = 2.75$ GeV 
and $a^{-1}(\beta=6.4) = 3.66$ GeV.

\noindent
Another way to  exhibit  
the scaling is to extract a $\Lambda$ parameter as a function of the scale 
and look for a plateau at high scale.
Like in \cite{luscher} we define  in general the $\Lambda$ parameter as~:
\bea
\Lambda \equiv &\mu&\exp\left (\frac{-2 \pi}
	{\tilde\beta_0
        \alpha_\mom(\mu^2)}\right)\times
        \left(\frac{\tilde\beta_0  \alpha_\mom(\mu^2)}
	{4 \pi}\right)^{-\frac {\tilde\beta_1}
        { \tilde\beta_0^2}}\nn\\	
	&&\times \exp\left\{ - \int_0^{\alpha_\mom(\mu^2)} {\rm d} \alpha 
	\left[ {1 \over \beta(\alpha)} + { 2 \pi\over \tilde\beta_0 \alpha^2} 
	         - { \tilde\beta_1\over \tilde\beta_0^2 \alpha } \right] \right\}
		 \label{general}
\eea

\noindent
If we consider the expansion of the \momt $\beta$ function  
truncated at three-loop, (\ref{general})  can be integrated    to give~:   
 \bea
 \Lambda_\mom \kern -2pt &=& \mu\exp\left\{\frac{-2 \pi}
	{\tilde\beta_0
        \alpha_\mom(\mu^2)}\right\}
        \left(\frac{\tilde\beta_0  \alpha_\mom(\mu^2)}
	{4 \pi}\right)^{-\frac {\tilde\beta_1}
        { \tilde\beta_0^2}}
 \left(1+\frac {\tilde\beta_1\ \alpha_\mom}{2\pi\tilde\beta_0}+
 \frac
 {\tilde\beta_2\ \alpha_\mom^2}
 {32\pi^2\tilde\beta_0}\right)^{\frac{\tilde\beta_1}{2\tilde\beta_0^2}}
 \nn\\
 & &\times \exp
\left\{\frac{\tilde\beta_0\tilde\beta_2-4\tilde\beta_1^2}
{2\tilde\beta_0^2\sqrt{\Delta}}\left[
\arctan\left(\frac{\sqrt{\Delta}}{2\tilde\beta_1+\tilde\beta_2
\alpha_\mom/4\pi}\right)
-\arctan\left(\frac{\sqrt{\Delta}}{2\tilde\beta_1}\right)\right]\right\}\, ,\nn\\
 \label{lambda3}
 \eea
 where $\Delta\equiv 2\tilde\beta_0\tilde\beta_2-4\tilde\beta_1^2$ 
 ($\Delta \ >\  0$ in our case). This can be consistently expanded
 to give:
 \bea
 \Lambda_\mom \simeq & & \mu \exp\left (\frac{-2 \pi}
	{\tilde\beta_0
        \alpha_\mom(\mu^2)}\right)\times
        \left(\frac{\tilde\beta_0  \alpha_\mom(\mu^2)}
	{4 \pi}\right)^{-\frac {\tilde\beta_1}
        { \tilde\beta_0^2}}
 \left(1+\frac {8 \tilde\beta_1^2\ - \tilde\beta_0\tilde\beta_2}{16\pi\tilde\beta_0^3} 
          \alpha_\mom \right)\nn\\
	  \label{referee}
 \eea

 \noindent
 For each value of the scale $\mu$, there is a corresponding $\alpha_\mom$ and 
 we can associate an effective $\Lambda_\mom$ parameter through 
 the formulas above. 
 This effective $\Lambda_\mom$ parameter  should become a constant at sufficiently high
 $\mu^2$ when  the  scaling settles. 
 It should be noted that (\ref{lambda3}) gives larger plateaux 
  than  (\ref{referee}). But independently of the formula used
 to extract $\Lambda$,   an  estimate for the systematic error 
 due to the influence of 
 higher-order terms on this determination is needed. 
 It can be obtained if we  add by hand a
  four-loop term in the beta function and vary its coefficient $\tilde\beta_3$ 
 over a reasonable interval.   
 Evaluation of these effects will be given below.

\begin{figure}
\begin{center}
\leavevmode
\vspace*{-4.0cm}
\epsfysize=11.5truecm\epsfbox[105 200 520 860]{lambda_MS_1.7_asym_sin_6.0_16.eps}
\epsfysize=11.5truecm\epsfbox[105 200 520 860]{lambda_MS_1.7_asym_sin_6.0_24.eps}
\\
\leavevmode
\epsfysize=11.5truecm\epsfbox[105 200 520 860]{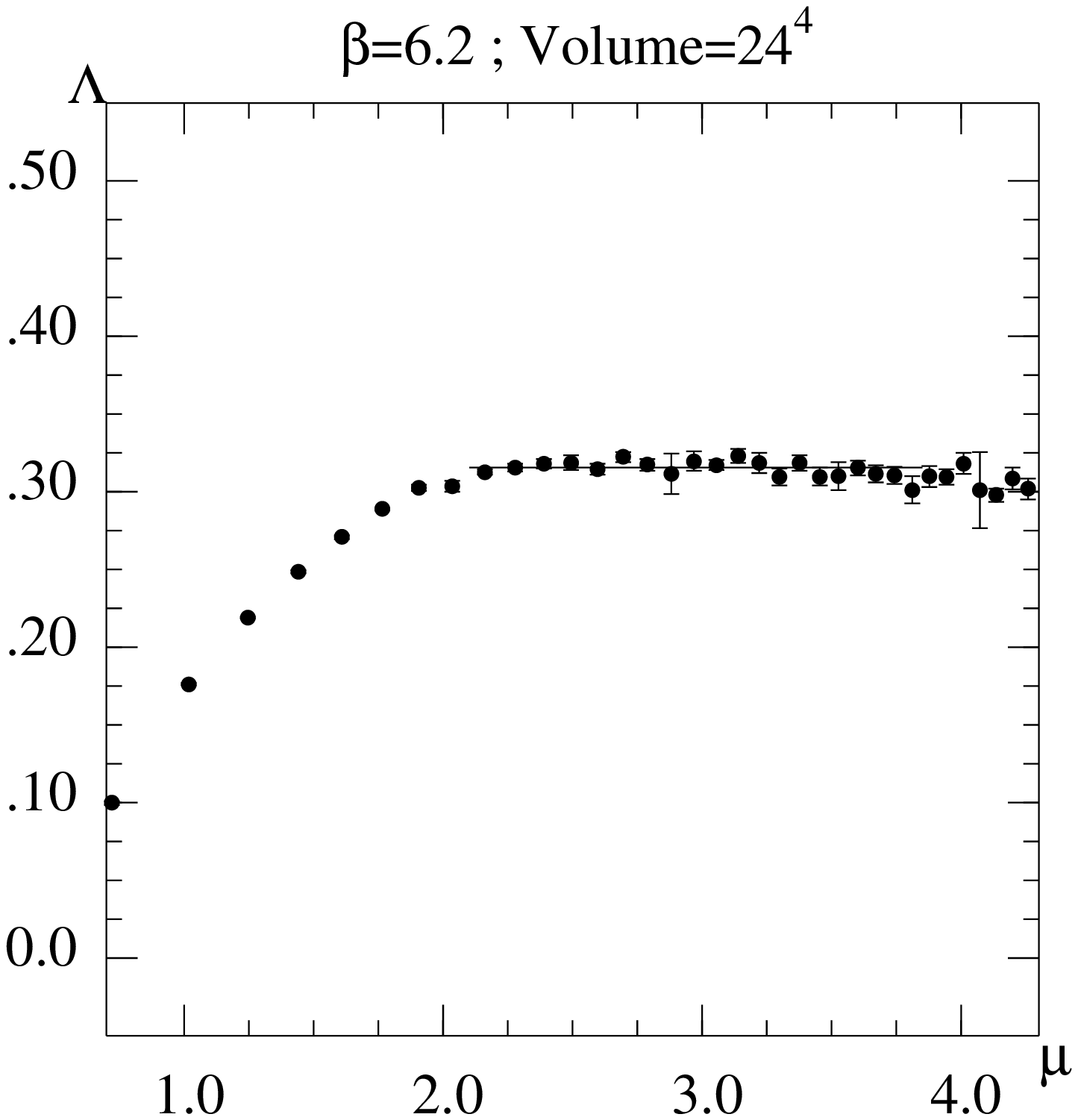}
\epsfysize=11.5truecm\epsfbox[105 200 520 860]{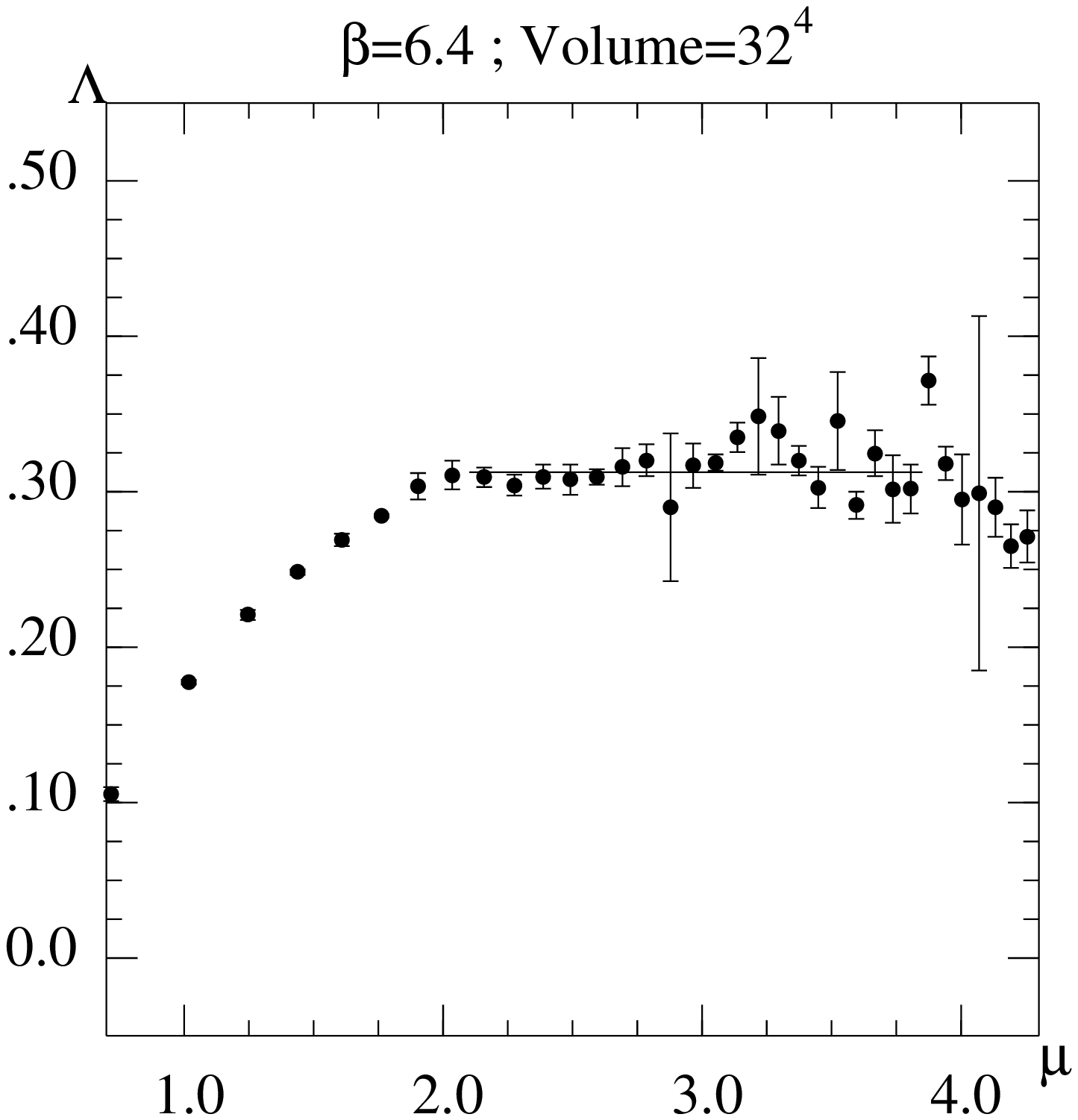}
\caption{$\Lambda_{\scriptstyle QCD}$ (in GeV) as a function of the scale $\mu$ (in GeV)
for different lattice spacings and volumes.}
\label{figure}
\end{center}
\end{figure}

 \noindent
 The asymptotic ratio $\Lambda_\mom \over \Lambda_\ms$ is known exactly \cite{alles}:
 $\Lambda_\ms \ = \ \exp(-\frac{70}{66}) 
 \times \Lambda_\mom$ $ \simeq\ 0.346 \times \Lambda_\mom$.   
 Let us  forget for the moment the influence of higher order terms and use
 the three-loop expression.
 In Fig.\ref{figure} we plot  the effective 
 $\Lambda_{\scriptstyle QCD} \equiv 0.346 \times\Lambda_\mom$ as a function of $\mu$
 for several values of the volume and  the lattice spacing.
 Scaling  manifests itself through the plateaux
 of $\Lambda_{\scriptstyle QCD}$ at large enough scale.
 The value of the plateau gives our "measurement" for  the flavorless $\Lambda_\ms$. 
 From our largest physical volume $(\beta,V)\ =\ (6.0,24^4)$
 (for which $V_{\rm phys}^{1/4} \simeq 2.4$ fm)
  we obtain~: 
$\Lambda_\ms= \left(303\pm 5\, {\rm MeV}\right)$.
For comparison we give  the  results 
from the   other lattices at $(\beta = 6.0 , V=16^4)$, 
$(\beta = 6.2 , V=24^4)$ and $(\beta = 6.4 , V=32^4)$, all three
with nearly the same physical volume, $V_{\rm phys}^{1/4} \simeq 1.7$ fm.
We found $\Lambda_\ms$ = 314(3), 313(4) and 312(9) respectively.
Scaling in $\beta$ is striking and a comparison between the two different
physical volumes shows that
finite volume effects   are  moderate ( a few \%).

\noindent
As explained previously, to estimate the systematic error from unknown higher order terms 
in the perturbative expansion we add a four-loop term in the beta-function 
and study the variation  of $\Lambda$ as a function of
the unknown coefficient $\tilde\beta_3$ varying up to  some $\tilde\beta_3^{lim}$. 
We conventionally choose to limit this interval 
 when the four-loop term in the beta function is equal in magnitude to the three-loop one
 at $\alpha = \alpha^{lim} \equiv   0.4$, 
 namely $\tilde\beta_3^{lim} = {2\pi\over\alpha^{lim}}\tilde\beta_2$.
We analyze the data and extract the  values for $\Lambda_\mom$ with
 \bea
 \Lambda_\mom \simeq & & \mu \exp\left (\frac{-2 \pi}
	{\tilde\beta_0
        \alpha_\mom(\mu^2)}\right)\times
        \left(\frac{\tilde\beta_0  \alpha_\mom(\mu^2)}
	{4 \pi}\right)^{-\frac {\tilde\beta_1}
        { \tilde\beta_0^2}}
 \left(1+\frac {8 \tilde\beta_1^2\ - \tilde\beta_0\tilde\beta_2}{16\pi\tilde\beta_0^3} 
          \alpha_\mom \right. \nn\\
	 && \left.+ \ \ {\alpha_\mom^2\over 2} 
	 \left( {2  \tilde\beta_0 \tilde\beta_1 \tilde\beta_2 - 8 \tilde\beta_1^3
	          -  \tilde\beta_0^2\tilde\beta_3 \over 32 \pi^2 \tilde\beta_0^4}
		  +\left(\frac {8 \tilde\beta_1^2\ - 
		      \tilde\beta_0\tilde\beta_2}{16\pi\tilde\beta_0^3}\right)^2 
	 \right)	      
  \right).
	  \label{referee4}
 \eea
From  the lattice with the largest physical volume this gives for our final result~:
\bea
 \Lambda_\ms= 295\, (5) (15)  \ 
 \frac {a^{-1}(\beta=6.0)}{1.97 {\rm GeV}}\, {\rm MeV}\quad \, .\label{final}
\eea
where the first error comes from the statistics
and the second one  from the systematics attached with the higher order terms.

\end{document}